\documentclass[twocolumn,apl,amsmath,amssymb,showpacs]{revtex4-1}
\usepackage{epsf}      
\usepackage{graphicx}
\usepackage{color}
\usepackage{gensymb}
\usepackage{amsmath}
\usepackage[outercaption]{sidecap}
\usepackage{multirow}
\begin{document}

\title{Evolution of complex magnetic phases and metal-insulator transition through Nb substitution in La$_{0.5}$Sr$_{0.5}$Co$_{1-x}$Nb$_x$O$_3$}
\author{Rishabh Shukla}
\affiliation{Department of Physics, Indian Institute of Technology Delhi, Hauz Khas, New Delhi-110016, India}
\author{R. S. Dhaka}
\email{rsdhaka@physics.iitd.ac.in}
\affiliation{Department of Physics, Indian Institute of Technology Delhi, Hauz Khas, New Delhi-110016, India}
\date{\today}

\begin{abstract}

We report the evolution of structural, magnetic, transport, and electronic properties of bulk polycrystalline La$_{0.5}$Sr$_{0.5}$Co$_{1-x}$Nb$_x$O$_3$ ($x =$ 0.025--0.25) samples. The Rietveld refinement of the x-ray diffraction patterns with R$\bar3$c space group reveals that the lattice parameters and rhombohedral distortion monotonously increase with the Nb$^{5+}$(4$d^0$) substitution ($x$). The magnetic susceptibility exhibits a decrease in the magnetic ordering temperature and net magnetization with $x$, which manifests that the Nb substitution dilutes the ferromagnetic (FM) double exchange interaction and enhances the antiferromagnetic (AFM) super-exchange interaction. Interestingly, for the $x>$ 0.1 samples the FM order is completely suppressed and the emergence of a glassy state is clearly evident. Moreover, the decrease in the coercivity (H$\rm_{C}$) and remanence (M$\rm_{r}$) with $x$ in the magnetic isotherms measured at 5~K further confirms the dominance of AFM interactions and reduction of FM volume fraction for the $x>$ 0.1 samples. More interestingly, we observe resistivity minima for the $x=$ 0.025 and 0.05 samples, which are analyzed using the quantum corrections in the conductivity, and found that the weak localization effect dominates over the renormalized electron-electron interactions in the 3D limit. Further, a semiconducting resistivity behavior is obtained for $x>$ 0.05, which follows the Arrhenius law at high temperatures ($\sim$160--320~K), and the 3D-variable range hopping prevails in the low-temperature region ($<$160~K). The core-level photoemission spectra confirm the valence state of constituent elements and the absence of Co$^{2+}$ is discernible.
\end{abstract}

\maketitle

\section{\noindent ~Introduction}

The exotic physical properties of LaCoO$_3$ are predominantly associated with the spin-state transition of Co$^{3+}$ \cite{RaccahPRB67, BarmanPRB94, AbbatePRB93, ZobelPRB02, ChainaniPRB92, ShuklaPRB18, SenarisJSSC95_2}, where it exhibits a nonmagnetic charge transfer type insulating ground state below $\sim$100~K and an insulator to metal transition near about 500~K \cite{AbbatePRB93, ZobelPRB02}. The ground state of LaCoO$_3$ is associated with the Co$^{3+}$ low-spin (LS) state (t$^6_{2g}$e$^0_g$), which evolves to a paramagnetic state (near 100~K) due to the emergence of high spin (HS; t$^4_{2g}$e$^2_g$) and/or intermediate spin (IS; t$^5_{2g}$e$^1_g$) states. Though the transition around 100~K was believed to be from LS to HS state \cite{AsaiPRB94, HaverkortPRL06, PodlesnyakPRL06}; however, this explanation was amended with the inclusion of the IS state, which was introduced in ref. \cite{KorotinPRB96} by LDA+U band structure calculations. At the same time, the evolution of the spin-state transition with external perturbation still remains enigmatic after several efforts, where a combination of the LS/IS scenario was supported in the explanation of insulator-to-metal transition near 500~K \cite{ZobelPRB02, KorotinPRB96}, which contradicts with the description given using the LS/HS combination \cite{RaccahPRB67, SenarisJSSC95_2}. Interestingly, the presence of Jahn-Teller distortion in LaCoO$_3$ strengthens the presence of IS state \cite{YamaguchiPRB97, ShuklaJPCC19, VankoPRB06, SatoJPSJ09}. Moreover, the IS state of Co$^{3+}$ is energetically near to the LS state (difference of $\sim$ 10~meV) and exists at lower energy as compared to the HS state, while in Co$^{4+}$ the HS state has a lower energy ($\sim$1~eV) with respect to the LS state \cite{KorotinPRB96, BarmanPRB94, ChainaniPRB92}. This energy difference ($\rm\Delta{E}$) between the spin-states of Co ions can be controlled by the volume of the governing CoO$_6$ octahedron via various external parameters like temperature and mechanical pressure \cite{VankoPRB06}, magnetic field \cite{SatoJPSJ09}, and chemical pressure \cite{PrakashJALCOM18, SenarisJSSC95_1, MutaJPSJ02}. This great tunability of the physical properties of cobaltites make them fascinating for  practical applications, like catalysis \cite{DiasCEC20}, solid oxide fuel cells \cite{HuangJES98}, sensors \cite{MuhumuzaIECR20}, batteries \cite{ZhangCS21, ChandrappaJMCA20}, etc.

Intriguingly, the substitution of divalent cations (like Ca, Sr, Ba) at La$^{3+}$ site results in the emergence of tetravalent Co$^{4+}$ ions, and a chemical pressure effect increases the lattice volume owing to the larger ionic radii of divalent cations, and a dramatic modification in the magnetic, and transport properties are observed \cite{PrakashJALCOM18, SenarisJSSC95_1, ShannonAC76, ImadaRMP98, ShannonAC76}. According to Imada {\it et al.}, the La$_{1-x}$Sr$_x$CoO$_3$ is a distinct filling-control system, where holes are solely responsible for the metallic conduction and ferromagnetism \cite{ImadaRMP98}. Also, the substitution of larger size Sr$^{2+}$ (1.44~\AA) as compared to La$^{3+}$ (1.36~\AA) enhances the crystal symmetry and unit cell volume, which are believed to stabilize the IS state of Co$^{3+}$ ions and do not change the structure up to $x=$ 0.55 \cite{PrakashJALCOM18, SenarisJSSC95_1, MineshigeJSSC96}. Moreover, the magnetic phase diagram of La$_{1-x}$Sr$_x$CoO$_3$ manifests a spin-glass state in 0.05$<x\le$0.18 and a ferromagnetic cluster glass for 0.18$\le x\le$0.50 due to the dominance of double-exchange FM interaction (Co$^{3+}$-O$^{2-}$-Co$^{4+}$) over the super-exchange (Co$^{4+}$-O$^{2-}$-Co$^{4+}$)/(Co$^{3+}$-O$^{2-}$-Co$^{3+}$) interactions \cite{SenarisJSSC95_1, ItohJPSJ94, KhanPRB12, WuPRL05, MiraPRB99, PhelanPRL06}. Also, an insulator to metal transition is induced for $x\ge$0.18 and further evolves to a complete metallic state for $x\ge$0.30 \cite{SenarisJSSC95_1, ItohJPSJ94}, and the magnetoelectronic phase separation is reported up to $x=$ 0.5 where FM metallic clusters are embedded in the AFM insulating matrix \cite{MutaJPSJ02, HePRB07}. More interestingly, the substitution of Nb$^{5+}$ (4$d^0$) in LaCo$_{1-x}$Nb$_x$O$_3$ converts Co$^{3+}$ into Co$^{2+}$ and induces structural transition owing to the larger ionic radius of Nb$^{5+}$ (0.64~\AA) and Co$^{2+}$ (0.75~\AA) as compared to Co$^{3+}$ (0.55~\AA) \cite{ShannonAC76, ShuklaPRB18}. The substitution of Nb$^{5+}$ induces the enhancement in the unit cell volume and stabilizes the HS state of Co$^{3+}$ and Co$^{2+}$ ions and drives the system towards a more insulating nature \cite{ShuklaPRB18}. Also, the perovskite cobaltites exhibit a strong coupling between the spin, charge, and lattice, where the correlation between charge carriers and localized spins plays a crucial role \cite{IvanovaPU09, LeeRMP85}, and can be tuned by cation substitution \cite{KrienerPRB04, SamalJPCM11}. Overall, it is convincing that the substitution of Sr at the La site and/or Nb at the Co site demonstrate interesting changes in the physical properties of LaCoO$_3$ (diamagnetic insulator) \cite{KumarPRB20, KumarPRB22}. However, to the best of our knowledge, a study to examine the effect of Nb substitution at the Co site in La$_{0.5}$Sr$_{0.5}$CoO$_3$ (ferromagnetic metal) has not been explored. 

Therefore, in this paper, we investigate the evolution of structural, magnetic, transport, and electronic properties of bulk La$_{0.5}$Sr$_{0.5}$Co$_{1-x}$Nb$_x$O$_3$ ($x =$ 0.025--0.25) samples. The Rietveld refinement of x-ray diffraction (XRD) patterns reveals the monotonous increase of unit cell parameters and rhombohedral distortion with $x$. The temperature-dependent susceptibility manifests the dilution of the magnetic order with $x$ due to nonmagnetic Nb$^{5+}$(4d$^0$), which dominates for $x>$0.1, and the appearance of a glassy state is evident. Moreover, the isothermal MH loops measured at 5~K manifest that coercivity (H$\rm_{C}$) and remanence (M$\rm_{r}$) decrease for $x>$0.1 due to the dominance of AFM interactions and reduction of FM volume fractions. More interestingly, we observe resistivity minima for the $x=$ 0.025, and 0.05 samples, which are analyzed using the quantum corrections in the conductivity, and find that the weak localization effect dominates over the renormalized electron-electron interactions in 3D limit. Moreover, a semiconducting behavior for the $x\ge$0.1 is analyzed with the Arrhenius model at high temperatures ($\sim$160--320~K) and 3D-variable range hopping conduction prevails in the low-temperature region ($<$160~K). Furthermore, the oxidation state of elements in La$_{0.5}$Sr$_{0.5}$Co$_{1-x}$Nb$_x$O$_3$ ($x=$ 0.1, 0.15, 0.25) is studied using core-level photoemission spectroscopy.

\section{\noindent ~Experimental}

The polycrystalline samples of La$_{0.5}$Sr$_{0.5}$Co$_{1-x}$Nb$_{x}$O$_3$ ($x =$ 0.025--0.25) are synthesized through the conventional solid-state route using stoichiometric initial ratio of as purchased powders (purity $>$99.95\%) of SrCoO$_3$, Co$_3$O$_4$, and Nb$_2$O$_5$ and dried La$_2$O$_3$ (900$\rm^o$C for 6 hrs). All the initial powders are grinded for 6-8 hrs using the agate mortar pestle and then obtained material is heated at 1000$\rm^o$C for 48 hrs to ensure homogeneous mixing. After the first heating, obtained mixture is reground for 4-6 hrs and cold-pressed into pellets for the final heating of 1200--1400$\rm^o$C for 36 hrs, which ensures the formation of a pure-phase compound \cite{ShuklaPRB18, ShuklaJPCC19}. The structural characterizations are performed at room temperature with the Panalytical X'pert Pro diffractometer using the Cu-K$_{\alpha}$ radiation ($\lambda$=1.5406~\AA). The magnetic measurements are performed using DynaCool Magnetic Property Measurement System from Quantum Design, USA. The temperature-dependent resistivity measurements were performed at the Physical Property Measurement System (PPMS) from Cryogenic Limited, UK, and the PPMS from Quantum Design, USA. The core-level x-ray photoemission spectra are recorded at room temperature using the AXIS Supra from the company Kratos Analytical limited, using a monochromatic Al-K$_\alpha$ (1486.6~eV) source having an overall energy resolution of $\sim$0.5~eV. We use the charge neutralizer during the measurement due to the insulating nature of the samples. The core-level spectra are deconvoluted and fitted with the Voigt-peak shapes using IgorPro software after subtracting the Tougaard inelastic background.

\section{\noindent ~Results and Discussion}

\begin{figure}[h]
\includegraphics[width=3.55in,height=5.75in]{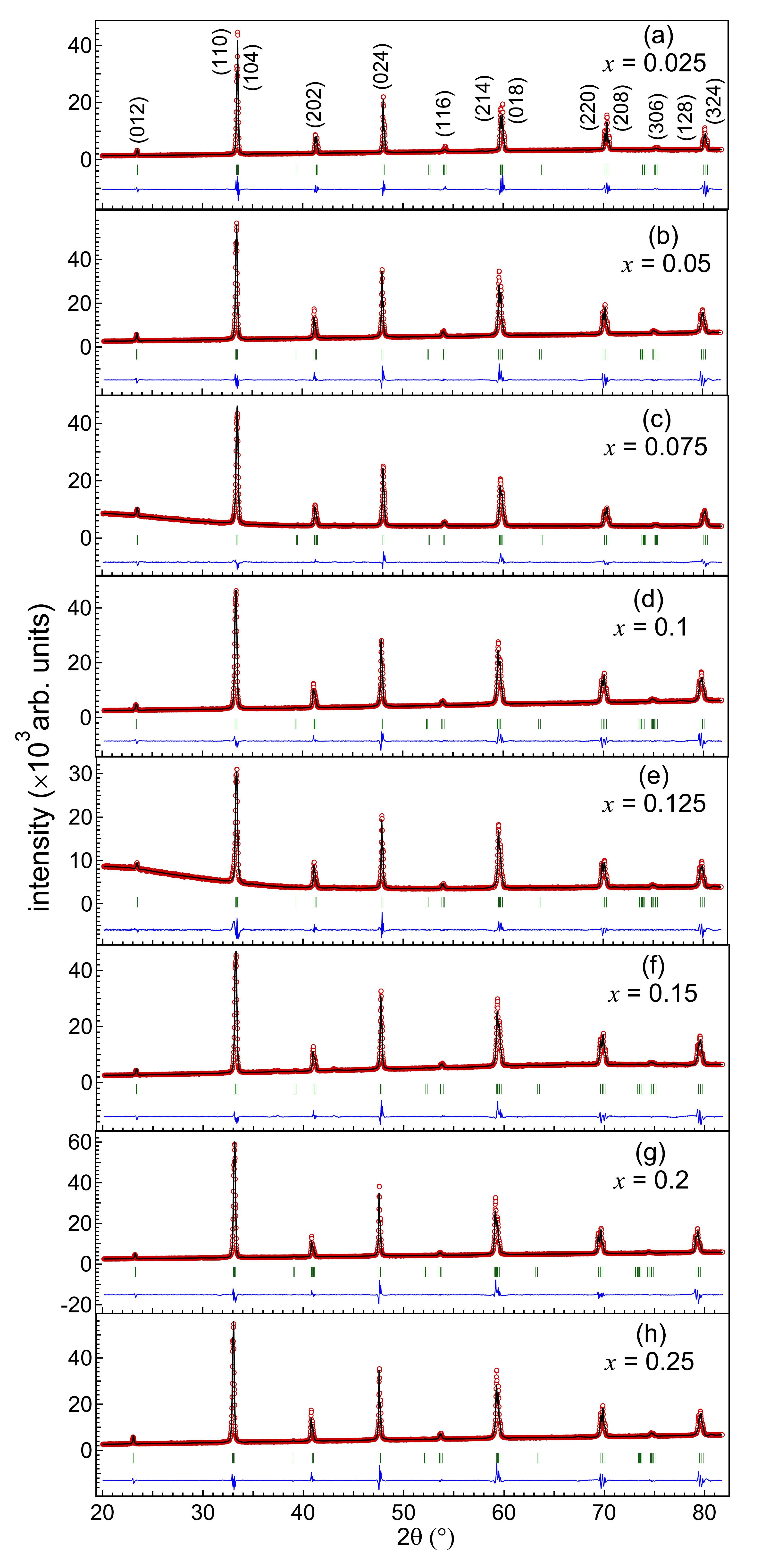}
\includegraphics[trim= 30 30 0 0,width=3.45in]{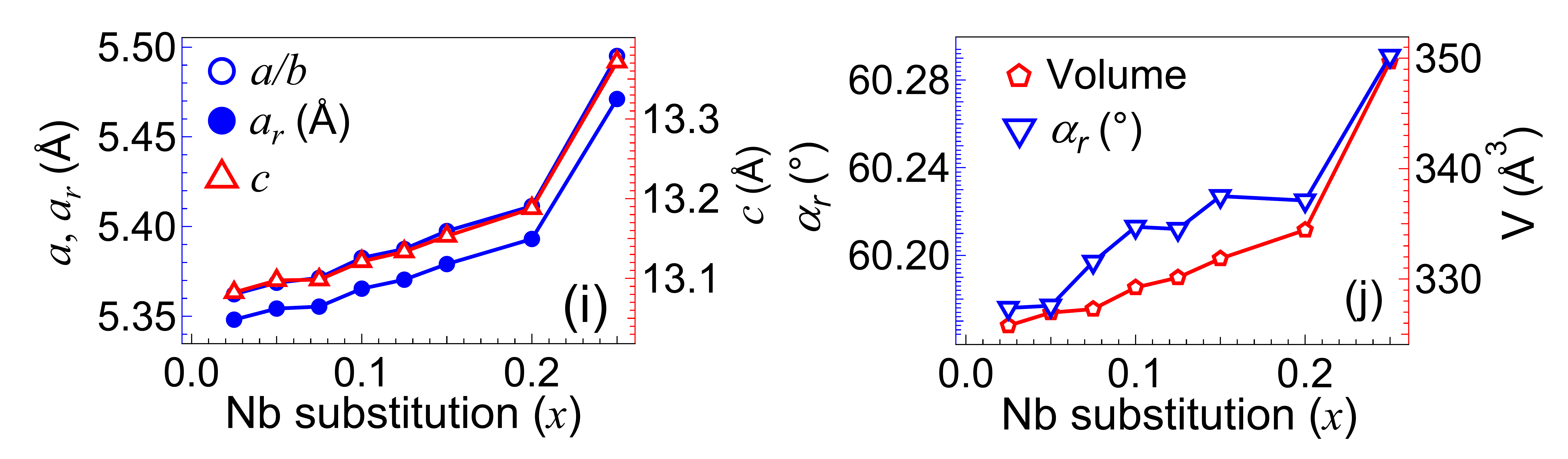}
\caption{(a--h) The Rietveld refined XRD patterns of La$_{0.5}$Sr$_{0.5}$Co$_{1-x}$Nb$_{x}$O$_3$ ($x=$ 0.025--0.25) samples; where open red circles, solid black and blue line represent the experimental, refined, and difference between experimental and refined patterns, while green vertical markers manifest the Bragg positions corresponding to the R$\bar{3}$c space group. (i, j) The Rietveld refined unit cell parameters plotted as a function of $x$.}
\label{fig:XRD}
\end{figure}

First, we perform the Rietveld refinement of measured XRD patterns of the La$_{0.5}$Sr$_{0.5}$Co$_{1-x}$Nb$_{x}$O$_3$, as presented in Figs.~\ref{fig:XRD}(a--h) for the $x=$ 0.025--0.25 samples, respectively, which confirm the single phase and rhombohedral space group (R$\bar3$c) in hexagonal setting ($a=b$, and $\gamma=$ 120$^{\rm o}$). The Rietveld refined lattice parameters are summarized in Table-I of \cite{SI} including the unit cell parameters determined for an equivalent rhombohedral cell (a$_r$ and $\alpha_r$) in the rhombohedral setting of the R$\bar3$c space group. Interestingly, the rhombohedral and hexagonal lattices are correlated via the symmetry transformation and, therefore, the Rietveld refinement using rhombohedral space group (R$\bar3$c) in a hexagonal setting will also result in the values of rhombohedral unit cell parameters, i.e., cell length (a$_r$) and distortion angle ($\alpha_r$). Further, the $x$--dependence of the unit cell parameters are plotted in Fig.~\ref{fig:XRD}(i); $a=b$ (open blue circles) and $a_r$ (solid blue circles) on the left-axis, and $c$ (open red triangles) on the right-axis. Here, we observe a monotonous enhancement in the values of these parameters with increasing the $x$. Moreover, we find an enhancement in the rhombohedral angle ($\alpha_r$) and unit cell volume (V) with the $x$, as shown on the left (open blue inverted triangles) and right (open red pentagons) axes of Fig.~\ref{fig:XRD}(j), respectively. This systematic monotonous enhancement in structural parameters with $x$ is accredited to the large ionic radii of Nb$^{5+}$ (0.64~\AA) as compared to the Co$^{3+}$ (0.545~\AA~in LS and 0.61~\AA~in HS) and Co$^{4+}$ (0.53~\AA~in HS) ions \cite{ShannonAC76}. Further, the enhancement in $\alpha_r$ manifests that the cationic substitution results in a higher degree of distortion of (Co/Nb)O$_6$ octahedra and therefore participates in the evolution of the spin-states of Co-ions with Nb substitution (discussed later) \cite{KumarJPCL22}. It has been reported that the substitution of Sr$^{2+}$ at the La site in LaCoO$_3$ suppresses the rhombohedral distortion ($\alpha_r$) and cubic crystal symmetry (Pm$\bar3$m) dominates for $x\ge$0.55 \cite{SenarisJSSC95_1, MineshigeJSSC96}, whereas the substitution of Nb$^{5+}$ at Co site drives the rhombohedral crystal symmetry from orthorhombic and then to monoclinic \cite{ShuklaPRB18}. 
Intriguingly, in the present case, the observation of an enhancement in the $\alpha_r$ with $x$ demonstrates a dominance of Nb$^{5+}$ ions on the structural parameters. Moreover, the observation of sharper diffraction peaks with much higher intensity (30--60k) indicates the absence of any microscopic inhomogeneities in these samples \cite{HaggertyJPCM04}.   

\begin{figure*}
\includegraphics[width=7.2in]{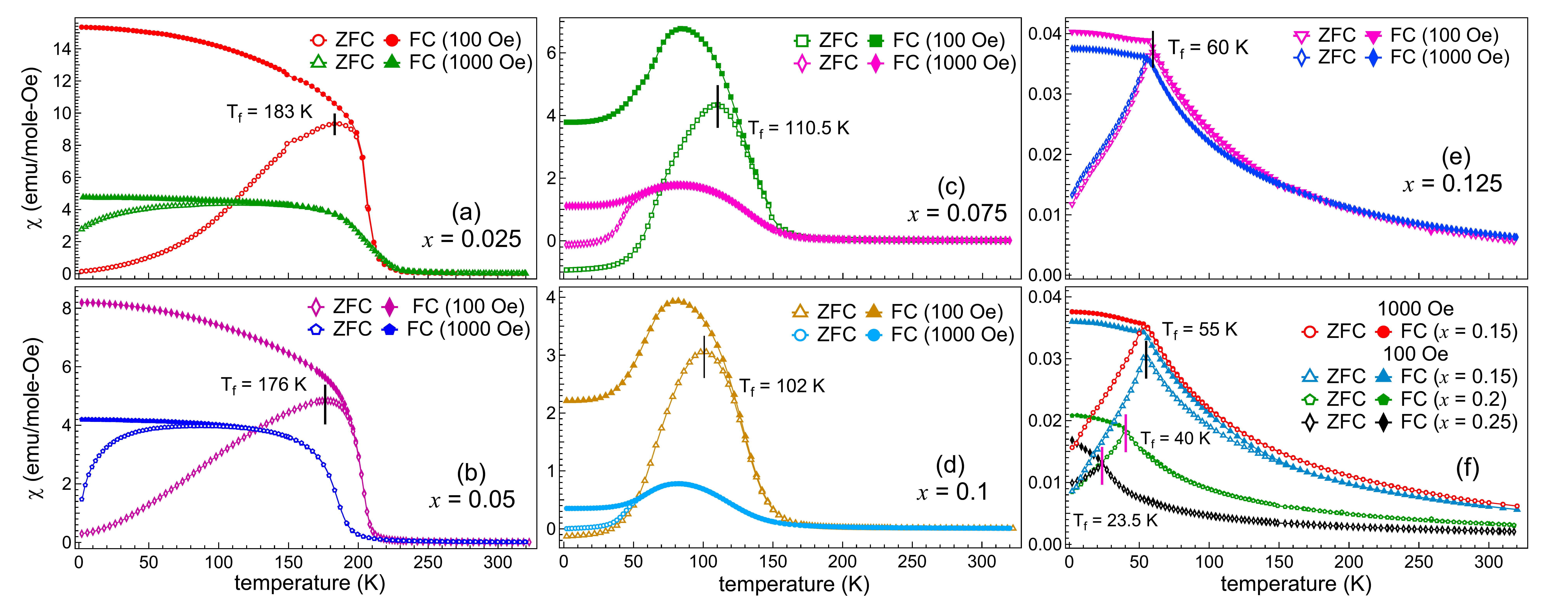}
\caption{A comparison of the temperature-dependent zero-field-cooled (ZFC) and field-cooled (FC) dc-magnetic susceptibility data recorded at 100~Oe and 1000~Oe applied magnetic field: (a) $x=$ 0.025, (b) $x=$ 0.05, (c) $x=$ 0.075, (d) $x=$ 0.1, and (e) $x=$ 0.125, and in (f) for the $x=$ 0.15 (100~Oe and 1000~Oe), $x=$ 0.2 (100~Oe), and $x=$ 0.25 (100~Oe) samples.}
\label{fig:ZFC+FC}
\end{figure*}

As discussed in the Introduction section, it is interesting to note here that the phenomenon of double exchange interaction in the Co$^{3+}$(IS)-O-Co$^{4+}$(LS) path is found to be responsible for the itinerant FM in La$_{0.5}$Sr$_{0.5}$CoO$_3$, which is achieved due to the spin pumping owing to the Sr$^{2+}$ substitution that converts Co$^{3+}$ ions into the Co$^{4+}$ ions and stabilizes the IS state of Co$^{3+}$ \cite{SenarisJSSC95_1, PrakashJALCOM18, ItohJPSJ94}. More interestingly, in the present case, the non-magnetic Nb$^{5+}$ (4$d^0$) substitution acts as a magnetic dilution in the FM state of La$_{0.5}$Sr$_{0.5}$CoO$_3$, and that suppresses the long-range FM ordering as Nb$^{5+}$ converts Co$^{4+}$ back to Co$^{3+}$ and hence reduces the double exchange mechanism. Also, in this case, the increased concentration of Co$^{3+}$ enhances the dominance of AFM super-exchange interaction through the Co$^{3+}$-O-Co$^{3+}$ path \cite{JonkarP53, BhidePRB75}. Therefore, in order to understand the expected complex magnetic behavior with Nb substitution in La$_{0.5}$Sr$_{0.5}$Co$_{1-x}$Nb$_{x}$O$_3$, we measure the temperature-dependent zero-field cooled (ZFC) and field-cooled (FC) magnetization at an applied magnetic fields of 100~Oe (for the $x=$ 0.025--0.25) and 1000~Oe (for the $x=$ 0.025--0.15), as shown in Figs.~\ref{fig:ZFC+FC}(a--f). Interestingly, we find a complex magnetic ordering below the transition temperature, as shown in Figs.~\ref{fig:ZFC+FC}(a--d), for $x=$ 0.025--0.1 samples, respectively and also highlighted in Figs.~1(a--f) of \cite{SI} by plotting the first-order-derivative of the dc-susceptibility. The sharp minima in Fig.~1(a) of \cite{SI} are clearly visible at 207~K and 204~K for the $x=$ 0.025 and 0.05 samples, respectively, measured at 100~Oe. Moreover, in Figs.~1(b, c) of \cite{SI}, the minima are observed at 203~K and 185~K when measured at 1000~Oe for the $x=$ 0.025 and 0.05 samples, respectively. Similarly, the minima are found to be at 135~K (100~Oe), and 131~K (1000~Oe) for the $x=$ 0.075 sample, and 130~K (100~Oe) and 121~K (1000~Oe) for the $x=$ 0.1 sample, as shown in Figs.~1(d--f) of \cite{SI}. 

It reveals that the magnetic ordering temperature is sensitive to the measuring field \cite{NamPRB99}, and we find that a small substitution of Nb$^{5+}$ (4$d^0$) significantly suppresses the magnetic ordering temperature (T$\rm_{MO}$) from 250--253~K for the $x=$ 0 sample \cite{HaggertyJPCM04, HoJMMM19} to 205$\pm$2~K for the $x=$ 0.025 sample, which may define the onset of magnetic dilution. Also, in Fig.~\ref{fig:ZFC+FC} (a), the maximum FC magnetic susceptibility decreases from $\sim$20~emu/mole-Oe ($x=$ 0) \cite{NamPRB99, HoJMMM19} to 15~emu/mole-Oe ($x=$ 0.025), and from $\sim$5.5~emu/mole-Oe ($x=$ 0) \cite{YoshiiPRB03, KhiemJAP05} to 4.8~emu/mole-Oe ($x=$ 0.025) when the measurements are done at 100~Oe and 1000~Oe, respectively. This type of complex magnetic ordering remains enigmatic in literature; for example, it is believed to be FM cluster glass \cite{SenarisJSSC95_1, ItohJPSJ94} and/or FM to paramagnetic transition \cite{MukherjeePRB00, WuPRL05}. Furthermore, in Figs.~\ref{fig:ZFC+FC}(a, b), the FC magnetization increases monotonously up to the lowest measured temperature (2~K), whereas the ZFC magnetization shows a cusp with broad peaks at around 183~K and 176~K (100~Oe) and 108~K and 88~K (1000~Oe) for the $x=$ 0.025 and 0.05 samples, respectively. It has been reported that the cusp in the ZFC magnetization can be associated with the anisotropy in the FM clusters available in the AFM matrix, and that is associated with the intra-cluster interaction and inter-cluster anisotropy in the matrix \cite{PodlesnyakPRB11}. Interestingly, a large bifurcation between the ZFC and FC magnetization further confirms the higher anisotropy in these samples \cite{SenarisJSSC95_1}. However, we find that this anisotropy decreases with the substitution of Nb$^{5+}$ ions and applied magnetic field, which plays a crucial role in the suppression of the intra-cluster FM interactions. 

\begin{figure*}
\includegraphics[width=7.2in]{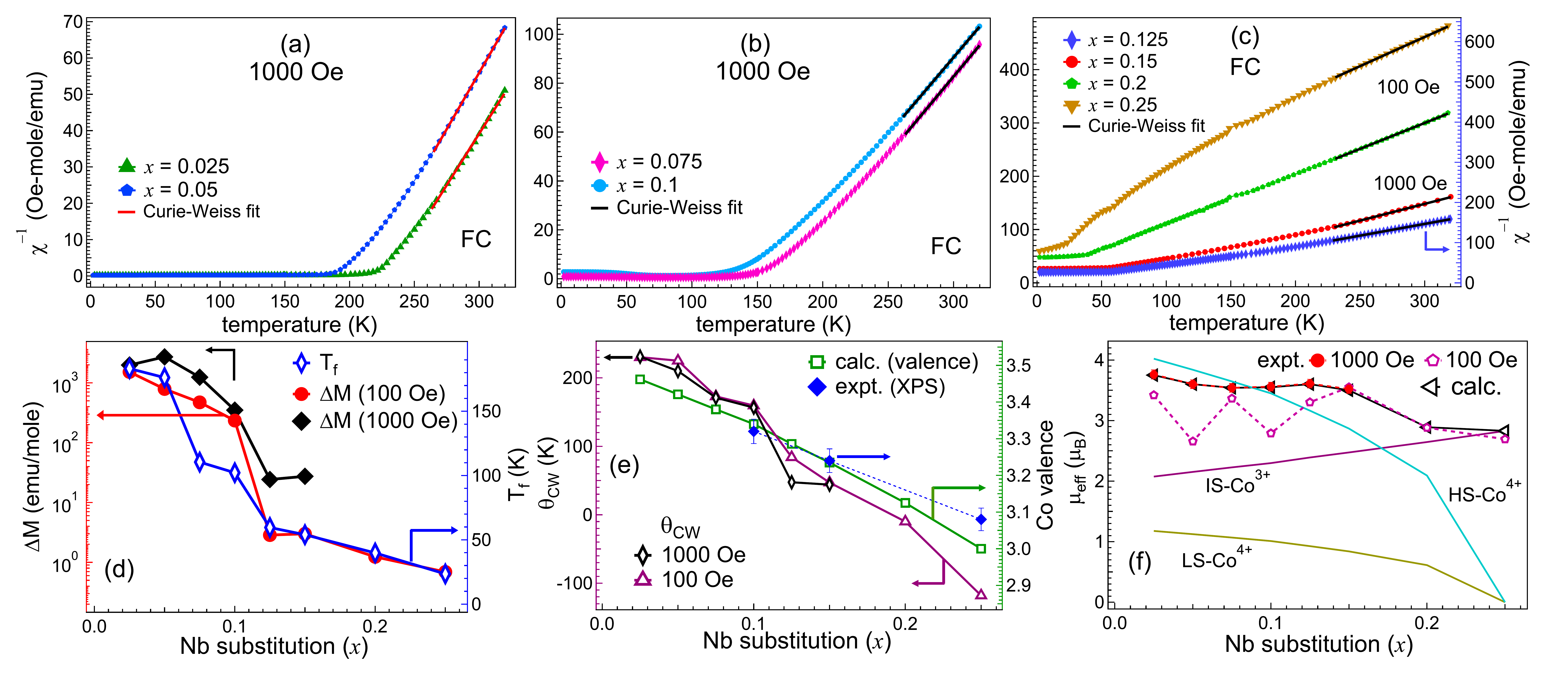}
\caption{A linear fit (solid black lines) of the inverse susceptibility data (recorded at 100~Oe and/or 1000~Oe) manifests the Curie-Weiss law in the paramagnetic region of (a) 260--320~K for the $x=$ 0.025 and 0.05 samples, (b) 260--320~K for the $x=$ 0.075 and 0.1 samples, (c) 230--320~K for the $x=$ 0.125--0.25 samples. (d) The thermomagnetic irreversibility, $\rm\Delta{M}$, for 100~Oe (solid red circles) and 1000~Oe (solid black rhombus) on a semilogarithmic left-axis, and peak/freezing (T$\rm_f$) temperature from the ZFC magnetization (open blue rhombus) on right-axis as a function of $x$, and (e) the Curie-Weiss temperature, $\rm\theta_{CW}$, for 100~Oe (open magenta triangles) and 1000~Oe (open black rhombus) along with the average cobalt valence, $n$ [open green rectangles (calculated) and solid blue rhombus (experimentally obtained using XPS)], and (f) a comparison of experimentally determined values of effective magnetic moments for susceptibility data recorded at 100~Oe and 1000~Oe magnetic fields and theoretically calculated values considering different possible spin-states of the Co ions.}
\label{fig:CW}
\end{figure*}

More importantly, we find a drastic change in the behavior of ZFC and FC magnetization curves for the $x=$ 0.075 and 0.1 samples, which exhibit a decreasing trend below a cusp and then remain constant to 2~K, as shown in Fig.~\ref{fig:ZFC+FC}(c, d). Here, the values of FC magnetic susceptibility at the peak are found to be around 6.8~emu/mole-Oe (100~Oe) and 1.8~~emu/mole-Oe (1000~Oe) for the $x=$ 0.075 sample, and 3.9~emu/mole-Oe (100~Oe) and 0.8~emu/mole-Oe (1000~Oe) for the $x=$ 0.1 sample. These values decrease below $\approx$83~K and almost saturates to around 3.8~emu/mole-Oe (100~Oe), 1.1~emu/mole-Oe (1000~Oe), and 2.2~emu/mole-Oe (100~Oe), 0.35~emu/mole-Oe (1000~Oe) at the lowest measured temperature (2~K) for the $x=$ 0.075 and 0.1 samples, respectively. The decreasing behavior of FC magnetic susceptibility is usually observed in the samples with an antiferromagnetic ordering \cite{AjayPRB20} and/or samples with a higher volume fraction of the AFM interactions as compared to the FM interactions \cite{TapatiPRB11, WangPRB99}. The magnetization values are found to be further decreasing with $x$, which is consistent with the fact that more Nb$^{5+}$ (4$d^0$) concentration acts like a magnetic dilution. Further, the difference between FC and ZFC magnetization ($\rm\Delta M$, thermomagnetic irreversibility) at the lowest temperature is correlated with the presence of FM and/or AFM volume fraction in the sample. A decrease in the value of $\rm\Delta M$ with $x$ is observed, i.e., 472~emu/mole (100~Oe), 1240~emu/mole (1000~Oe) for the $x=$ 0.075, and 234~emu/mole (100~Oe), 350~emu/mole (1000~Oe) for the $x=$ 0.1 sample, see the left-axis [solid red circles (100~Oe) and solid black rhombus (1000~Oe)] of Fig.~\ref{fig:CW}(d). This suggests that we are restricted to collect the magnetization from the smaller volume fraction of FM and higher fraction of AFM phase with increasing $x$ \cite{TapatiPRB11}. The ZFC magnetic susceptibility values are found to be negative at 2~K when measured at 100~Oe magnetic field, which increased significantly when measured at 1000~Oe for the $x=$ 0.075 and 0.1 samples [see Figs.~\ref{fig:ZFC+FC}(c, d)].  Here, a large magnetic anisotropy due to the strong competitive AFM and FM interactions might be responsible for the field/temperature induced magnetization reversal behavior \cite{RenNature98, SarkarAPL13}, which is also consistent with the variation of $\rm\Delta M$ with the applied magnetic field. Interestingly, we observe a further decrease in the T$\rm_{MO}$ to 135~K (100~Oe), 131~K (1000~Oe) for the $x=$ 0.075 sample, and 130~K (100~Oe), 121~K (1000~Oe) for the $x=$ 0.1, and an increase in the broadening of the transition temperature region with $x$ can be attributed to the enhanced magnetic disorder in the material and dilution of the FM double exchange interaction \cite{FerrariPM11, SahuJMMM06, GennesPRB60}. Moreover, in Figs.~\ref{fig:ZFC+FC}(c, d), the cusp/peak in the FC and ZFC magnetization at 100~Oe is observed at different temperatures like 83.5~K and 110.5~K ($x=$ 0.075) and 82~K and 102~K ($x=$ 0.1), respectively. The corresponding values at 1000~Oe are found to be 82~K and 84~K ($x=$ 0.075), and 80~K and 81.5~K ($x=$ 0.1), respectively.  The peak/cusp in the FC curves appears at a lower temperature as compared to the ZFC, which is due to the difference in the measurement protocols as the FC measures the higher FM volume fraction compared to the ZFC protocol \cite{TapatiPRB11}. Therefore, confirmation of the decreased FM volume fraction in higher $x$ samples is obtained from a smaller difference in the cusp/peak value of FC and ZFC magnetization, i.e., $\rm\Delta{T_f}$=27~K (100~Oe), 2~K (1000~Oe) for the $x=$ 0.075 sample, and 20~K (100~Oe), 1.5~K (1000~Oe) for the $x=$ 0.1 sample. Furthermore, the ZFC and FC susceptibility remains constant from $\sim$35~K (100~Oe) and 28~K (1000~Oe) to the lowest measured temperature for both the FC and ZFC protocols [see Fig.~\ref{fig:ZFC+FC}(c, d)], and this is expected to be a state of magnetic glass where the competitive FM and AFM magnetic clusters are frozen arbitrarily in the sample at low temperatures \cite{TapatiPRB11}.

Interestingly, for the $x=$ 0.125--0.25 samples we find a substantial change in the magnetic behavior, as shown in Fig.~\ref{fig:ZFC+FC}(e, f), respectively. For example, the FM ordering is suppressed due to reduction in the concentration of Co$^{4+}$ ions and the evolution of AFM interactions due to an enhancement in Co$^{3+}$, which may result in the evolution of a glassy state \cite{DuPRB07, BinderRMP86} for the $x \ge$ 0.125 samples. The magnetic susceptibility value at 2~K dropped abruptly to ~0.04~emu/mole-Oe for the $x=$ 0.125 and 0.15 samples, which indicates that the magnetic order is very much sensitive to the percolation limit in the La$_{0.5}$Sr$_{0.5}$Co$_{1-x}$Nb$_x$O$_3$ samples. Moreover, we can see that all samples in the 0.125 $\le~x\le$ 0.25 range manifest analogous behavior of FC and ZFC magnetization curves with temperature [see Fig.~\ref{fig:ZFC+FC}(e, f)], where both the FC and ZFC magnetization increase monotonously with lowering the temperature, and after a peak/cusp in ZFC (known as the freezing temperature, T$\rm_f$), the FC magnetization further increases at a lower rate while the ZFC magnetization starts decreasing up to the lowest measured temperature (2~K). Additionally, we observe that the freezing temperature T$\rm_f$ (peak/cusp in ZFC) decreases with $x$, see right-axis (open blue rhombus) of Fig.~\ref{fig:CW}(d), such that it has the value of 58$\pm2$~K ($x=$ 0.125), $\approx$55~K ($x=$ 0.15), 40~K ($x=$ 0.2), and 23.5~K ($x=$ 0.25). Also, the value of $\rm\Delta{M}$ measured at 100~Oe decreases from 3~emu/mole for the $x=$ 0.125 to 0.7~emu/mole for the $x=$ 0.25 sample, as shown on the left-axis (solid red circles) of Fig.~\ref{fig:CW}(d), which further suggests a minimal volume fraction of the FM magnetic phase in these samples as discussed above. Therefore, we can conclude that the substitution of Nb$^{5+}$ (4$d^0$) plays a crucial role in the dilution of the FM order and inducing a magnetic glassy state at low temperatures \cite{DuPRB07, BinderRMP86, MukherjeePRB96}.

In order to determine the effective magnetic moment ($\rm\mu_{eff}$), we plot the inverse FC magnetic susceptibility in Fig.~\ref{fig:CW}(a--c) for the $x=$ 0.025--0.15 (1000~Oe) and the $x=$ 0.2, 0.25 (100~Oe) samples. Moreover, the inverse susceptibility data for the $x=$ 0.025--0.15 samples measured at 100~Oe and 1000~Oe are compared in the Figs.~2(a--f) of \cite{SI}. We fit the inverse susceptibility data using the Curie-Weiss (C-W) equation [$\chi^{-1} = (T-\theta_{\rm CW}$)/C], where $\rm C$ is C-W constant and $\rm\theta_{\rm CW}$ is C-W temperature in the paramagnetic region for all the La$_{0.5}$Sr$_{0.5}$Co$_{1-x}$Nb$_x$O$_3$ samples. The obtained slope in the C-W fit, for the $x=$ 0.025--0.15 (1000~Oe) and $x=$ 0.2, 0.25 (100~Oe) samples, is used to determine the $\rm\mu_{eff}$ and the intercept is utilized to estimate the $\rm\theta_{\rm CW}$, which are summarised in Table II of \cite{SI}. Further, we plot the $x$--dependence of $\rm\mu_{eff}$ (solid red circles for 1000~Oe and open magenta pentagon for 100~Oe) in Fig.~\ref{fig:CW}(e), which shows a monotonous decrease in the values as a function of $x$. Additionally, we observe a monotonous decrease in the $\rm\theta_{CW}$ values (magenta open triangles, left axis) as well as the calculated average Co valence, $n$ (green open rectangles) with $x$, as shown in Fig.~\ref{fig:ZFC+FC}(h). The value of $n$ is calculated using the general formula of La$^{3+}_{0.5}$Sr$^{2+}_{0.5}$Co$^n_{1-x}$Nb$^{5+}_x$O$^{2-}_3$, where $x$ is the percentage of Nb substitution. Here, the experimentally obtained values of the Co valence state (solid blue rhombus) from the core-level photoemission spectra of Co 2$p$ are also compared (discussed later), which matches well for the $x=$ 0.1 and 0.15 samples, while slightly deviates for the $x=$ 0.25 sample. Note that, the decreasing trend of these parameters manifests an overall evolution of the AFM interactions over FM interactions and reduction of the magnetic anisotropy with $x$ \cite{JoyJPCM98}. Moreover, we calculate the theoretical value of $\rm\mu_{eff}^{calc.}$ using the equation, $\mu_{\rm eff}^{\rm calc.} = \sqrt{(1-p)[\mu_{\rm eff}^{4+}]^2+p[\mu_{\rm eff}^{3+}]^2}$, where $p$ is the fraction of the Co$^{3+}$ ions, and the $\rm\mu_{eff}^{4+}$ and $\rm\mu_{eff}^{3+}$ correspond to the 4+ and 3+ valence states of Co, respectively. These theoretically calculated values of $\rm\mu_{eff}^{calc.}$ considering different spin-states of Co$^{3+}$ and Co$^{4+}$ ions. For the $x=$ 0 sample, Xu {\it et al.} reported that the Co$^{3+}$ ions present in the IS state, whereas the Co$^{4+}$ ions stabilize in a mixture of LS and HS states \cite{XuPLA06}. Similarly, we calculate the $\rm\mu_{eff}$ values considering composition weighted spin-states of Co$^{3+}$/Co$^{4+}$ ions in La$_{0.5}$Sr$_{0.5}$Co$_{1-x}$Nb$_x$O$_3$, which are plotted in Fig.~\ref{fig:CW}(f) as a function of $x$. The calculated values of $\rm\mu_{eff}$ are found to be in good agreement with the experimental $\rm\mu_{eff}^{expt.}$ values, as shown in Fig.~\ref{fig:CW}(f). All the extracted parameters are summarized in Table II of \cite{SI}.

\begin{figure}[h]
\includegraphics[width=3.6in]{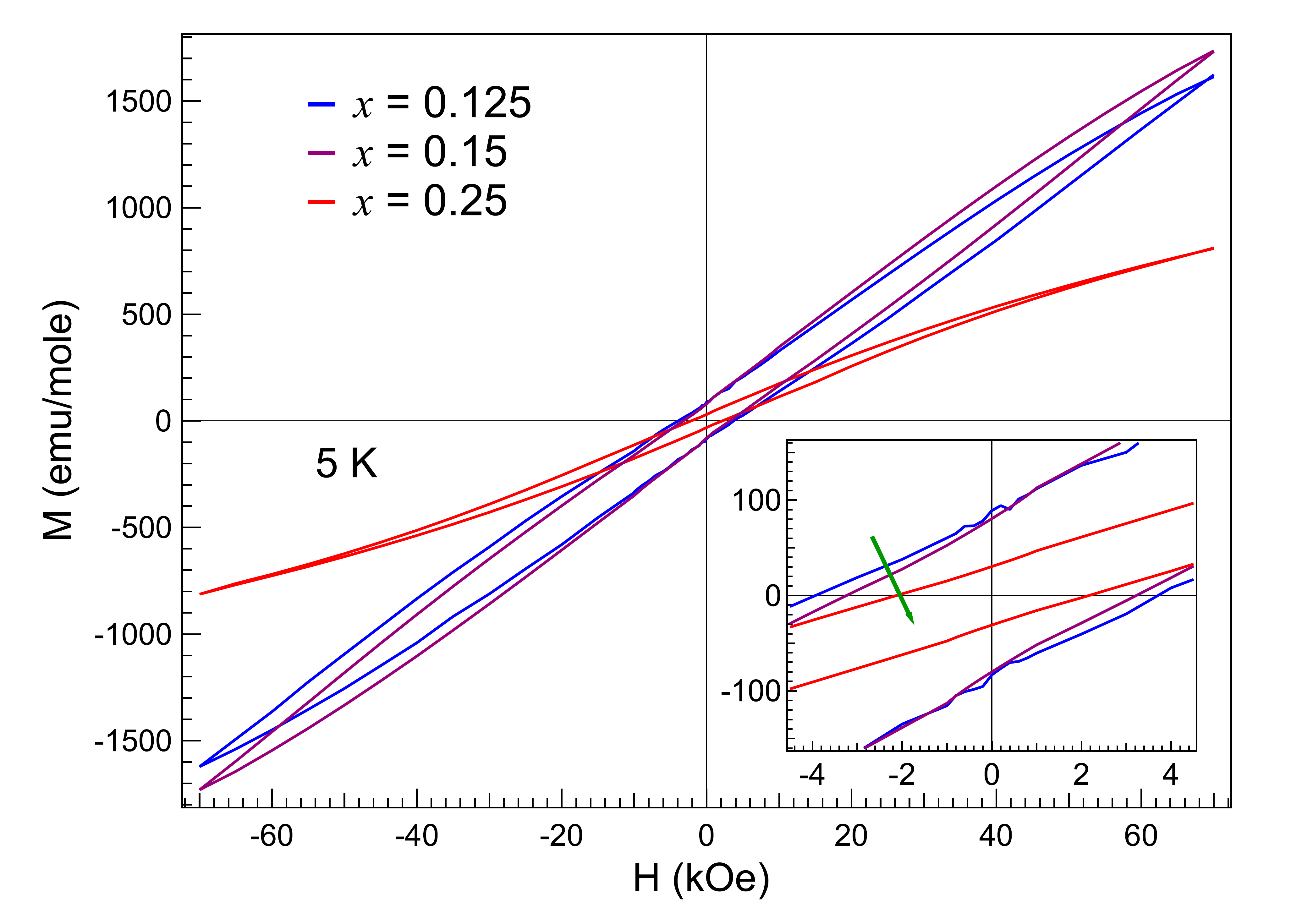}
\caption{The isothermal M-H loops recorded at 5~K within the applied magnetic field of $\pm$70~kOe for the $x=$ 0.125, 0.15, and 0.25 samples; the inset highlights the variation of magnetization near the zero-magnetic field.}
\label{fig:MH}
\end{figure}

Furthermore, to understand the magnetic interaction in the highly competitive FM and AFM regions, we measure the isothermal MH loops for the $x=$ 0.125, 0.15, and 0.25 samples at 5~K within $\pm$70~kOe applied magnetic field under the ZFC protocol and shown in Fig.~\ref{fig:MH}. Interestingly, a non-saturating symmetric behavior observed in the magnetization is due to the constant increase of spin-reorientation in the field direction. The estimated values of coercivity ($\rm H_C$) are found to be 3.8, 3.2, and 2.2~kOe, and the remanence ($\rm M_r$) values are 86, 80, and 31~emu/mole for the $x=$ 0.125, 0.15, and 0.25 samples, respectively. We find that the values of H$\rm_{C}$ and M$\rm_{r}$ decrease with increasing the $x$, as highlighted with a green arrow in the inset of Fig.~\ref{fig:MH}, which manifests that the higher Nb$^{5+}$ (4$d^0$) concentration suppresses the volume fraction of the spin-pinning boundaries as well as the non-homogeneity of the magnetic phase \cite{KhanPRB12}. Moreover, the virgin isothermal curves measured at 5~K deviate from the linear behavior and show an upturn at higher fields (convex nature), see Fig.~3(a) of \cite{SI}, which suggests a metamagnetic nature for the $x=$ 0.125 and 0.15 samples. A more closer look indicates two critical magnetic fields at $\approx$16 (H$\rm_{m_1}$) and 35~kOe (H$\rm_{m_2}$) (arrows in Figs.~3(b, c) of \cite{SI}) \cite{PandeyPRB20}. These can be associated with competitive AFM and FM interactions at low fields; however, a linear increase of magnetization beyond H$\rm_{m_2}$ suggests that AFM interactions dominate at higher fields \cite{ElovaaraJPCM12, PandeyPRB20}. However, for the $x=$ 0.25 sample, the saturating nature of the M-H curve at high magnetic fields ($>$ 40~kOe) manifests the field-induced spin-reorientation \cite{MajumdarJPCM13}.

\begin{figure}[h]
\includegraphics[width=3.5in]{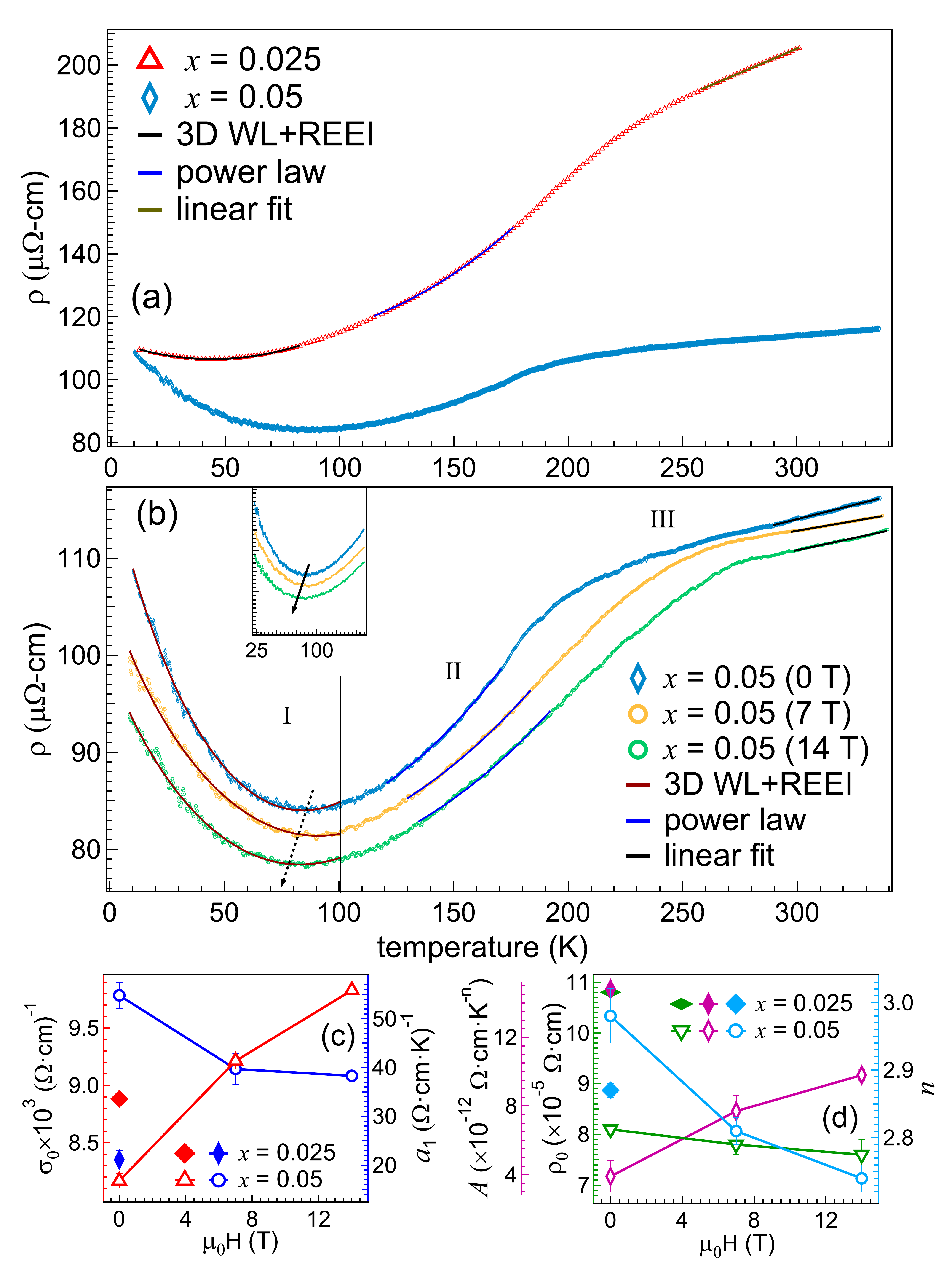}
\caption{(a) A comparison of the temperature-dependent resistivity of the $x=$ 0.025 and 0.05 samples at zero magnetic field, (b) the variation in resistivity with the applied magnetic field (0, 7, and 14~T) for the $x=$ 0.05 sample, experimental and fitted data are shown in the open symbols and solid lines, respectively. The inset in (b) highlights the decrease in the resistivity minimum (T$\rm_{min}$) with the magnetic field. (c) The variation of the low-temperature (below 100~K) fitting parameters using equation (1), $\sigma_0$ (left-axis) and a$_1$ (right-axis), and (d) the variation in fitting parameters in 120--200~K range using equation (3), $A$ (first-left-axis), $\rho_0$ (second-left-axis), and power law-exponent ($n$) with the magnetic field (for the $x=$ 0.05 sample) and for both the $x=$ 0.025 and 0.05 samples at the zero-magnetic field.}
\label{fig:RT1}
\end{figure}

To investigate the temperature dependent transport behavior in La$_{0.5}$Sr$_{0.5}$Co$_{1-x}$Nb$_x$O$_3$, first the $\rho$--T curves are compared in Fig.~\ref{fig:RT1}(a) for the $x=$ 0.025 and 0.05 samples. The resistivity varies linearly above T$\rm_{MO}$$\sim$250~K, whereas a gradual change in the slope, i.e., a subtle decrease in the resistivity below T$\rm_{MO}$ can be correlated to the conversion of Co$^{3+}$ ions from HS state to IS state in this temperature range. The changes can be attributed to the reduction in the scattering because of the spin-disorder owing to the FM ordering in the sample below T$\rm_{MO}$ \cite{MahendiranPRB96}. Interestingly, we observe a minimum (T$\rm_{min}$) near 77~K and 83~K for the $x=$ 0.025 and 0.05 samples, respectively, as shown in  Fig.~\ref{fig:RT1}(a). The resistivity increases below T$\rm_{min}$, which is the signature of metal to insulator transition (T$\rm_{MI}$), and can be associated with the spin-state transition from the IS to the LS state \cite{YamaguchiPRB97}. The metallic nature in the $x=$ 0.025 and 0.05 samples can be understood due to the presence of a larger volume fraction of Co$^{4+}$ ions, as in La$_{1-x}$Sr$_x$CoO$_3$ for $x\ge0.18$ \cite{SenarisJSSC95_1}. Interestingly, a similar resistivity behavior is reported near T$\sim$100~K in magnetic oxides \cite{RoyAP95} and is believed to be a disorder-induced effect \cite{LeeRMP85} as well as this temperature scales with the spin-state transition of Co ions from HS to IS \cite{ZobelPRB02, ShuklaPRB18}. Further, at low temperatures, the resistivity starts increasing and that indicates the transition of Co$^{3+}$ ions to the LS state, as reported for LaCoO$_3$ \cite{ZobelPRB02, ShuklaPRB18}. In Fig.~\ref {fig:RT1}(b) we compare the $\rho$-T behavior of $x=$ 0.05 sample measured at 0, 7, and 14~Tesla applied magnetic fields. The applied magnetic field enhances the exchange interaction strength and the observed negative magnetoresistance behavior below $\approx$100~K can be associated with the enhanced density of states near the Fermi level \cite{ChangPRL09}. Intriguingly, in the quantum transport regime at low temperatures, the phenomenon of weak localization (WL) or weak antilocalization (WAL) correction in the conductivity arises due to the interference of the electron waves traveling back and forth in the material and is termed as the quantum corrections in the conductivity (QCC) \cite{LeeRMP85}. The negative magnetoresistance in the $x=$ 0.05 sample manifests the dominance of WL as the field suppresses the magnetic disorder and hence the resistivity \cite{LeeRMP85}. Also, the interference of the two different electron waves gives rise to the renormalized electron-electron interaction (REEI) \cite{LeeRMP85}. Therefore, a least-square minimization method is used to fit the experimental data by considering the contributions of WL and REEI in both three-dimensional (3D) and two-dimensional (2D) limits.

The $\rho$-T behavior can be approximated by the following equations (1) and (2) for the 3D and 2D limits of WL and REEI effects, respectively \cite{LeeRMP85, HerranzPRB05}:
\begin{eqnarray}
 \rho(T) = \frac{1}{(\sigma_0 + a_1T^{p/2} + a_2T^{1/2})} + b^{\prime}T^2
\end{eqnarray}
\begin{eqnarray}
\rho(T) = \frac{1}{\sigma_0 + a_3lnT} + b^{\prime}T^2	
\end{eqnarray}
where $\sigma_0$ is the residual conductivity (1/$\rho_0$), $a_1T^{p/2}$ accounts for the 3D WL contributions, and $a_2T^{1/2}$ is attributed to the 3D REEI corrections, and $a_3$lnT is accredited to the combined contributions of the WL and REEI in the 2D limit. The variable $p$ is the index of localization effects in the system and can be influenced by different types of scattering mechanisms. The value of $p=$ 2 implies the dominance of electron-electron scattering, whereas $p=$ 3 is attributed to electron-phonon scattering \cite{LeeRMP85}. The last term ($b^{\prime}T^2$) accounts for the Boltzmann term and implies the classical low-temperature dependence of resistivity withholding the Matthiessen rule \cite{HerranzPRB08}. We use equation (1) to fit the experimental data using a combination of 3D WL and REEI at low temperatures (below 100~K), as shown by the solid lines in Figs.~\ref {fig:RT1}(a, b). Moreover, in Figs.~2(c, d) of \cite{SI} we show a comparison of the residual resistivity, $\Delta\rho(\%) = [\rho_{\rm fit}(T)-\rho(T)]/\rho(T)\times$100\%) by fitting the zero field data considering the WL plus REEI corrections in both 3D (solid red circles) and 2D (solid blue triangles) limits for the $x=$ 0.025 and 0.05 samples, respectively. This clearly shows a larger deviation with respect to the zero line in case of the 2D limit as compared to the 3D limit, which confirms the dominance of the 3D limit of WL plus REEI corrections. Now to understand the origin of the quantum effects whether from WL or REEI \cite{HerranzPRB05, LeeRMP85}, we measure the resistivity of the $x=$ 0.05 sample at high applied magnetic fields of 7 and 14~Tesla and present in Fig.~\ref {fig:RT1}(b). We find that a higher magnetic field reduces the resistivity and shifts the minimum point slightly towards the low temperatures [as shown in the inset of Fig.~\ref {fig:RT1}(b)]. This suggests for diminishing of the QCC effects, which can be attributed to the suppression of wave coherence of the electron under an applied magnetic field and gives a strong evidence of the WL correction at low temperatures because the REEI correction should be independent of the magnetic field \cite{LeeRMP85}. Further, in Fig.~\ref {fig:RT1}(b), we compare the fitting [using equation (1)] of resistivity data measured at 0, 7, and 14~Tesla. The obtained values of $\sigma_0$ (open red triangles) and a$_1$ (open blue circles) from the fitting at low temperatures ($<$100~K) are plotted in Fig.~\ref{fig:RT1}(c) on the left and right axes, respectively. A significant increase in $\sigma_0$ can be correlated to the negative magnetoresistance and a decrease in the value of $a_1$ is attributed to the reduction of the quantum corrections with the magnetic field and manifests the dominance of WL effects. The parameters obtained from the low-temperature fit for the $x =$ 0.025 and 0.05 samples are summarized in Table III of \cite{SI}. 

In order to analyze the resistivity behavior in the intermediate temperature range (120--200~K), we use the power law to fit the least square minimization to the experimental data, see Figs.~\ref{fig:RT1}(a, b):
\begin{eqnarray}
\rho(T) = \rho_0 + AT^n
\end{eqnarray}
where $\rho_0$ is the residual resistivity, $n$, and $A$ are the fitting parameters to be utilized to quantify the disorder in the FM oxides \cite{LeeRMP85}. In the present case, the $\rho_0$ values are estimated in the range 8--11~$\times$10$^{-5}$~$\Omega$-cm for both the samples (see Table III of \cite{SI}). The values of coefficient $A=$ 0.4--1.5$\times$10$^{-11}$~$\Omega$-cm-K$^{-2}$ are found to be comparable to the values obtained for the elemental ferromagnets, which are attributed due to electron-electron scattering \cite{VolkenshteinPSSB73}. Also, in ferromagnetic metal oxides, $n$ values can be used to quantify the disorder; for example, Herranz {\it et al.} reported the $n$ value change from $\sim$ 1.1 to 3 for the weak and strong disorder systems, respectively \cite{HerranzPRB08}. In our case, the $n$ value is found to be 2.9$\pm$0.05 for both samples at zero magnetic field, which suggests the presence of strong disorder in the system. This is also consistent with the observed non-linearity in $\rho$-T behavior and the co-existence of FM ordering and metal-insulator transition in these samples \cite{LeeRMP85}. Moreover, the fitting parameters $A$ (open magenta rhombus), $\rho_0$ (open green triangles), and $n$ (open cyan circles) are plotted on the right and left axes of Fig.~\ref {fig:RT1}(d), respectively. The exponent $n$ decreases from 2.95 (0~T) to 2.75 (14~T), which manifests a decrease in the disorder with the applied magnetic field \cite{HerranzPRB08}. The fitting parameters obtained using equation (2) are included in Table III of \cite{SI} for the $x=$ 0.025 and 0.05 samples. Further, the analysis of the $\rho$-T behavior of $x=$ 0.025 and 0.05 samples [Figs.~\ref{fig:RT1}(a, b)] in the high-temperature range (T$>$T$\rm_{MO}$) is important to study the effect of disorder on the electron-phonon coupling; therefore, we use the electronic transport theory based on Bloch-Gr$\rm\ddot{u}$neisen law, as given below \cite{GurvitchPRL78, HerranzPRB08}: 
\begin{eqnarray}
\begin{aligned}
\rho&=\rho_0+\rho_m +\frac{2\pi k_B}{\hbar e^2(n/m_{eff})}G(\Theta_D/T)\lambda'T\\
&=\rho_0+\rho_m+\gamma(T)T
\end{aligned}
\end{eqnarray}
where $\rho_0$, $\rho_m$, and $\gamma(T)$ are the residual resistivity, magnetic resistivity due to the spin scattering of the electrons (considered constant in paramagnetic phase at T$>$T$\rm_{MO}$), and $\gamma(T)$ is the temperature-dependent Gr$\rm\ddot{u}$neisen parameter, respectively. Here, $\gamma(T)$ is considered equal to [2$\pi k_B/\hbar e^2(n/m_{\rm eff})]G(\Theta_D/T)\lambda'$, where $n$ is the carrier density, m$_{\rm eff}$ is the effective mass of carriers, $G(\Theta_D/T)$ is the Gr$\rm\ddot{u}$neisen function. The values of $\gamma(T)$ are determined from the slope of the temperature-dependent linear fit of $\rho(T)$ in the high-temperature range \cite{HerranzPRB08}, see Figs.~\ref{fig:RT1}(a, b). The obtained fitting parameters are presented in Table III of \cite{SI}, where the $\gamma(T)$ values are found to be 31$\times$10$^{-8}$~$\ohm$-cm-K$^{-1}$ for the $x=$ 0.025, which is much higher as compared to the $x=$ 0.05 sample 6$\times$10$^{-8}$~$\ohm$-cm-K$^{-1}$. This manifests a higher degree of disorder in the $x =$ 0.025 sample. Further, for the $x=$ 0.05 sample, there is no significant change in the values of fitting parameters [$\rho_0$, $\rho_m$ and $\gamma(T)$] with the applied magnetic field, which suggests invariance of disorder at high temperatures. 

\begin{figure}[h]
\includegraphics[width=3.48in]{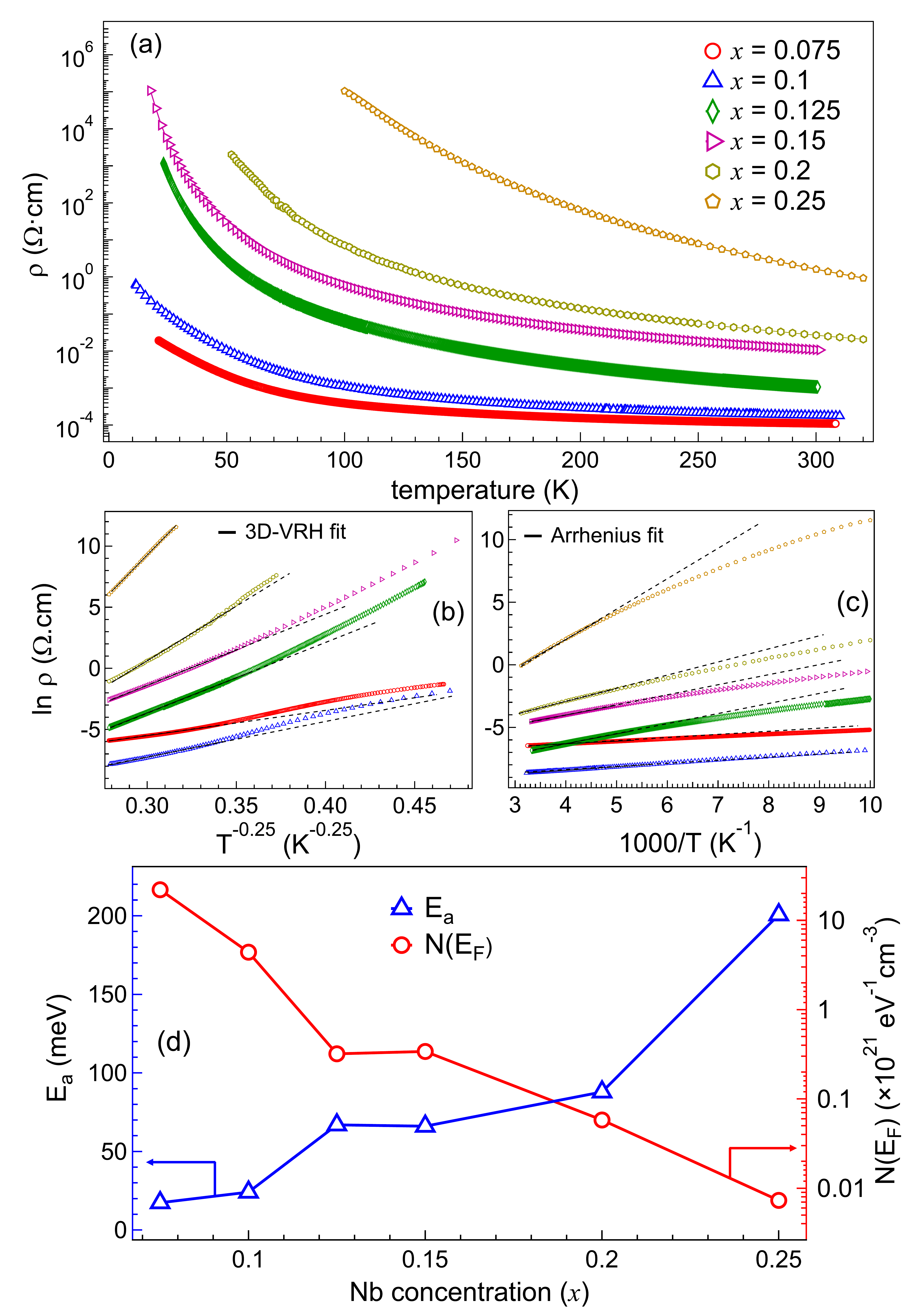}
\caption{(a) A comparison of the temperature-dependent resistivity variation of $x=$ 0.075--0.25 samples on the semi-logarithmic scale, (b) the low-temperature resistivity fitted with the 3D-VRH model, and (c) the high-temperature resistivity data fitted with the Arrhenius model; a solid-black line presents the fit for the corresponding model and the dotted lines mark the range where the experimental curve deviates from the respective fits, and (d) the activation energy (E$\rm_a$) extracted from the Arrhenius model on a linear scale and density of states near the Fermi level [N(E$\rm_F$)] extracted from the 3D-VRH model on a semi-logarithmic scale.}
\label{fig:RT2}
\end{figure}

In Fig.~\ref{fig:RT2}(a), we present the temperature dependent resistivity data (on a semilogarithmic scale) of the 0.075 $\le x \le$ 0.25 samples, which clearly show an increasing trend with decreasing temperature, i.e., a semiconducting/insulating nature. Interestingly, the resistivity increases monotonously with $x$ due to higher Nb$^{5+}$ concentration, which predominantly favors the insulating nature of these samples. This observation further confirms that the substitution of Nb$^{5+}$ converts metallic Co$^{4+}$ ions into Co$^{3+}$, which restricts the metallic conduction pathways. Moreover, the Nb$^{5+}$ does not participate in conduction due to the unavailability of conduction electrons in 4$d^0$. In order to understand the possibility of carrier conduction mechanisms in these samples, we analyze the experimental data in two different regions, the high-temperature (320--200~K) with the Arrhenius model (where charge carriers move through the band-conduction) and the low-temperature region (160--60~K) with the three-dimensional variable range hopping (3D-VRH) model (where charge carriers move between the ions through nearest-neighbor hopping in the delocalized states). The Arrhenius model can be expressed as \cite{ShuklaPRB18}, $\rho(T) = \rho_0 exp(E_a/k_BT)$, where, $\rho_0$ and E$_a$ are the exponential pre-factor and activation energy, respectively. In Fig.~\ref{fig:RT2}(c), we show the ln($\rho$) with 1000/T, and the slope of linear fit (solid black line) of these curves is utilized to determine the activation energy for each sample. The obtained values of activation energy are presented on the left-axis of Fig.~\ref{fig:RT2}(d) in a linear scale, where we obtain a monotonous enhancement of the activation energy, which is consistent with the dominance of insulating nature with $x$. Interestingly, a deviation of the experimental data from the Arrhenius model towards low temperatures (highlighted with the dotted black line) demonstrates a possibility of a different conduction mechanism \cite{AjayPRB20}. Therefore, we follow the three-dimensional variable-range hopping (3D-VRH) model, which can be given as \cite{ShuklaPRB18}, $\rho(T) = \rho_0exp\left(\frac{T_0}{T}\right)^{1/4}$, where, $\rho_0$ and T$_0$ are the exponential factors and characteristic temperatures, respectively. In Fig.~\ref{fig:RT2}(b), the best fit of the plot between ln($\rho$) and T$^{-1/4}$ validates the 3D-VRH model in the low temperature range, and the dotted black lines are shown where a deviation in the experimental data from the 3D-VRH model is visible. The slope of the linear fit is used to determine the characteristic temperature (T$_0$) and that is used to calculate the density of states (DOS) near the Fermi level [N(E$\rm_F$)] using the formula, N(E${\rm_F}$)=18/k$\rm_B$T$_0$$\lambda^3$, where $\lambda$ is the localization length of the conduction path. Moreover, the conducting path for the VRH is chosen here (Co/Nb)--O--(Co/Nb) and its value is determined from the Rietveld refinement for the calculation of the DOS. We show the variation in DOS on the right-axis of Fig.~\ref{fig:RT2}(d) in a semi-logarithmic scale, which manifests a monotonous decrease in N(E$\rm_F$) with $x$. Further, for the confirmation of the VRH model, we estimate the values of mean hopping distance (R$\rm_H$) and means hopping energy (W$\rm_H$), as given below \cite{PaulPRL73},
\begin{eqnarray}
R_{\rm H}&=&[9\lambda/8 \pi kTN(E_F)]^{1/4}~nm \\
W_{\rm H}&=& [3/4\pi R_H^3 N(E_F)]~eV
\end{eqnarray}
The calculated values of R$\rm_H$ and W$\rm_H$ are found to increase with $x$, as presented in Table III of \cite{SI}. The obtained parameters for the $x \ge$0.075 samples fulfill Mott's criteria for VRH conduction, $\lambda^{-1}$R$\rm_H\textgreater\textgreater$ 1 and W$\rm_H\textgreater\textgreater$ k$\rm_B$T, at low temperatures below 160~K. These results manifest the magnetic dilution due to Nb$^{5+}$ (4d$^0$) substitution, which leads to the reduction in the conduction pathways for the electron hopping and drives the system toward a more insulating state. We also note here that the temperature-dependent resistivity analysis includes the effect of the scattering of charge carriers at the grain boundary due to the polycrystalline nature of these samples \cite{ReissPRL86, SagdeoJMMM06}. 

\begin{figure}[h]
\includegraphics[width=3.5in]{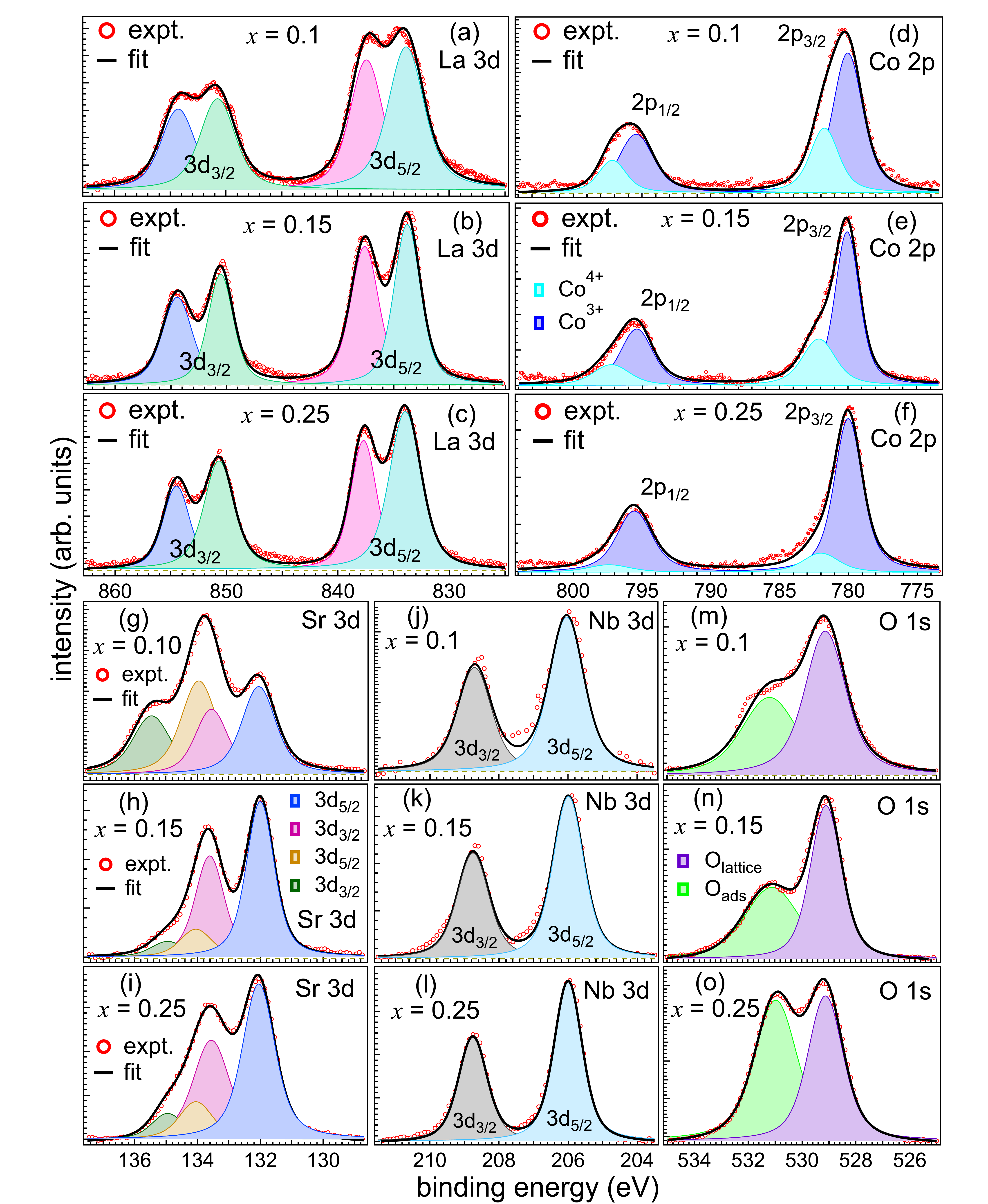}
\caption{The core-level x-ray photoemission spectra measured at room temperature and the deconvoluted components of (a--c) La 3$d$, (d--f) Co 2$p$, (g--i) Sr 3$d$, (j--l) Nb 3$d$, and (m--o) O 1$s$ for the $x=$ 0.1, 0.15 and 0.25 samples.}
\label{fig:XPS}
\end{figure}

Finally, we investigate the electronic properties of the $x=$ 0.1, 0.15, and 0.25 samples using core-level photoemission spectroscopy. In Figs.~\ref{fig:XPS}(a--c), we present the core-level La 3$d$ spectra, which show two peaks in each spin-orbit component of 3$d_{5/2}$ and 3$d_{3/2}$ having a separation $\approx$3.8~eV due to the transfer of the electron to the La 4$f^0$ from the oxygen valence band \cite{PrakashJALCOM18, LamPRB80}. Here, the binding energy (BE) values for the main components are 850.6~eV (3$d_{3/2}$) and 833.8~eV (3$d_{5/2}$) having a separation of 16.8~eV confirm the presence of La$^{3+}$ ions \cite{PrakashJALCOM18, LamPRB80}. More importantly, the spin-orbit split components 2$p_{3/2}$ and 2$p_{1/2}$ of Co 2$p$ core-levels are shown in Figs.~\ref{fig:XPS}(d--f). The deconvolution of Co 2$p_{3/2}$ with two peaks corresponds to the 3+ and 4+ oxidation states, where the BE of Co$^{3+}$ ions is 780~eV (2$p_{3/2}$) and 795.5~eV (2$p_{1/2}$), wheres the Co$^{4+}$ is found to be at 782.0~eV (2$p_{3/2}$) and 797.4~eV (2$p_{1/2}$), both having a separation of $\approx$15.5~eV, which are in good agreement with the refs. \cite{AjayJAP20, ChungSS76}. There is no signature of any satellite feature at 6~eV from the main peak of Co 2$p$ core-levels, which confirms the absence of Co$^{2+}$ \cite{ChungSS76}) in the $x=$ 0.1, 0.15 and 0.25 samples, see in Figs.~\ref{fig:XPS}(d--f). These results also conclude that the substitution of Nb$^{5+}$ converts Co$^{4+}$ ions into Co$^{3+}$ ions \cite{ShuklaJPCC19, ShuklaJPCC21}. Further, we determine the average valence ($n$) of the Co ions from the area ratio of the deconvoluted components and found it 68\% (Co$^{3+}$) and 32\%(Co$^{4+}$) for the $x=$ 0.1 sample, and 72\% (Co$^{3+}$) and 28\%(Co$^{4+}$) for the $x=$ 0.15 sample, which manifests an average valence of 3.32 ($x=$ 0.1) and 3.28 ($x=$ 0.15), and are found to be in good agreement with theoretically determined values of 3.34 and 3.24, respectively. For the $x=$ 0.25 sample, the ratio is 88\% (Co$^{3+}$) and 12\%(Co$^{4+}$) having an average valence of 3.1, which is slightly higher than the theoretically calculated value of 3. Further, in Figs.~\ref{fig:XPS}(g--i) we present the Sr 3$d$ core levels, which are deconvoluted with four components. The two strong components at the lower BE side are observed at 132~eV (3$d_{5/2}$) and 133.6~eV (3$d_{3/2}$) with a separation of 1.6~eV, which are in good agreement with the values reported for the Sr$^{2+}$ valence state \cite{AjayJAP20, SosulnikovJESRP92}. The other two weaker components at around 134.0~eV (3$d_{5/2}$) and 135.3 (3$d_{3/2}$) are from SrCO$_3$ traces present on the surface of the sample \cite{SosulnikovJESRP92}. Moreover, the Nb 3$d$ core-levels are presented in Figs.~\ref{fig:XPS}(j--l) for the $x=$ 0.1, 0.15 and 0.25 samples, respectively, where the deconvoluted components at 206~eV (3$d_{5/2}$) and 208.8~eV (3$d_{3/2}$) with 2.8~eV separation. These values are in good agreement with the refs.~\cite{ShuklaPRB18, AjayJAP20} and confirm the 5+ oxidation state of Nb in these samples. The O 1$s$ core-level spectra in Figs.~\ref{fig:XPS}(m--o) are deconvoluted into two peaks at the BE values of 529~eV and 531.1~eV, which correspond to the contributions from the lattice oxygen and surface adsorbed oxygen, respectively \cite{ImamuraJPCB00}.

\section{\noindent ~Conclusions}

In conclusion, we have investigated the evolution of structural, magnetic, and transport properties of bulk La$_{0.5}$Sr$_{0.5}$Co$_{1-x}$Nb$_x$O$_3$ ($x=$ 0.025--0.25). The Rietveld refinement of the x-ray diffraction patterns with R$\bar3$c space group, reveals that the lattice parameters and rhombohedral distortion monotonously increase with the increased Nb concentration. The magnetic susceptibility data manifest that the Nb substitution dilutes the double exchange interaction and a decrease in the magnetic ordering temperature and net magnetization are observed with $x$. Interestingly, for the $x>$0.1 samples, the magnetic interactions are dominated by the superexchange antiferromagnetic interactions. Moreover, the isothermal MH loops recorded at 5~K for the $x>$0.1 samples manifest that coercivity (H$\rm_{C}$) and remanence (M$\rm_{r}$) decrease with $x$ due to the dominance of AFM interactions and reduction of FM volume fractions. More interestingly, we observe a minimum in resistivity for the $x=$ 0.025 and 0.05 samples, which are analyzed using the quantum corrections in the conductivity where the weak localization effect dominates over the renormalized electron-electron interactions in the 3D limit. Moreover, a semiconducting behavior is observed for the $x>$ 0.05 samples, and the resistivity monotonously increases with higher Nb$^{5+}$(4$d^0$) concentration. This semiconducting resistivity behavior is analyzed with the Arrhenius model in the high-temperature ($\sim$160--320~K) and 3D-variable range hopping conduction in the low-temperature region ($<$160~K). Further, the core-level photoemission spectra of La 3$d$, Sr 3$d$, Co 2$p$, Nb 3$d$, and O 1$s$ confirm the valence state of constituent elements and affirms the absence of Co$^{2+}$, which are in accordance with the magnetization results. 

\section{\noindent ~Acknowledgements}

RS acknowledges DST, INSPIRE for the fellowship, and IIT Delhi for providing research facilities: XRD and PPMS at the Department of Physics, and PPMS and XPS at CRF. RS thanks Mr. Ramcharan Meena for his help in the transport measurements. RSD acknowledges SERB--DST for the financial support through a core research grant (project reference no. CRG/2020/003436).

\end{document}